\documentclass[11pt]{article}

\begin{document}

\title{
V.M. Red'kov\footnote{redkov@dragon.bas-net.by}\\
Dirac-K\"{a}hler equation in curved space-time, \\ relation between spinor and
tensor formulations\footnote{Translated version of a paper: VINITI
7.08.89, no 5336 - B89, Minsk,  1989;  Chapter 3   in: V.M. Red'kov,
Fields in Riemannian space  and the Lorentz group (in Russian).
Publishing House "Belarusian Science", Minsk, 2009.}\\[3mm]
{\small B.I. Stepanov Institute of physics, National Academy of
Sciences of Belarus} }

%\date{}

\maketitle

\begin{quotation}

A common view is that generalization of a wave equation on Riemannian space-time is substantially
 determined  by what a particle is -- boson or fermion.
As a rule, they say that tensor equations for bosons are extended in a simpler way then spinor equations
for fermions.
In that context, a very interesting   problem is  of extension a wave equation for
Dirac--K\"{a}hler field (Ivanenko--Landau field was  historically first term, also the term
a vector field  of general type was used).

The article relates a generally covariant tensor formalism to
a spinor one when these both are applied to description  of
the Dirac-K\"ahler   field in a Rimannian space-time.
Both methods are taken to be equivalent and the tensor equations are
derived from spinor ones. It is shown that, for characterization of
Dirac-K\"ahler's tensor components, two alternative approaches are
suitable: these  are whether a tetrad-based  pseudo tensor
classification or a generally coordinate pseudo tensor one.
By  imposing
definite restrictions on  the the Dirac-K\"ahler function,
we  have  produced  the  general
covariant form of wave  equations  for  scalar,  pseudoscalar,  vector,  and  pseudovector
particles.

\end{quotation}

\section{ Introduction}

Mathematical description of the concept of  elementary  particles
as certain relativistically invariant objects was found in the frames
of 4-dimensional Minkowski space-time. It is assumed that for any particle
there are given definite transformation properties of a corresponding  field and a wave equation
to  which that field obeys; wave equation must be Lorentz (or Poincar\'{e}) invariant:
Wigner \cite{1939-Wigner}, Pauli  \cite{1941-Pauli}, Bhabha
\cite{1949-Bhabha}, Harish-Chandra \cite{1947-Yarish-Chandra},
Gel'fand -- Yaglom   \cite{1948-Gel'fand}, Corson  \cite{1953-Corson}, Umezawa  \cite{1956-Umezawa},
 Shirokov  \cite{1957-Shirokov}, Bogush --  Moroz  \cite{1968-Bogush-Moroz},
 Fedorov в \cite{1979-Fedorov}).

 A common view is that generalization of a wave equation on Riemannian space-time is substantially
 determined  by what a particle is -- boson or fermion.
As a rule, they say that tensor equations for bosons are extended in a simpler way then spinor equations
for fermions. This believing evidently correlates with the fact: concepts of both flat  and curved  space model
are based on the notion of a vector.

In that context, a very interesting   problem is  of extension a wave equation for
Dirac--K\"{a}hler field (there are used other  terms as well: Ivanenko---Landau field, or a vector field  of general type).

Scientific literature  consecrated with this  field is enormous, it started
early in the development of quantum mechanical wave equations theory, just after
the concept of a particle with spin 1/2 arises.
In particular, news objects themselves,  spinors,  seemed  mysterious and obscure
in comparison with familiarized tensors.

The main feature of the Ivanenko---Landau field \cite{1928-Ivanenko}
was that it seemingly gave possibility to perform  smoothly transition from tensors to spinors,
in a sense it was an attempt to eliminate spinors  at all.
Different aspect of that relation were investigated by many authors:
Иваненко,  Ландау   \cite{1928-Ivanenko},
Lanczos  \cite{1929-Lanczos(1), 1929-Lanczos(2)},
 Juvet  \cite{1930-Juvet, 1932-Juvet},
 Einstein -- Mayer   \cite{1932-Einstein(1), 1933-Einstein(1),
1933-Einstein(2), 1934-Einstein},
 Frenkel \cite{1935-Frenkel}, Whittaker  \cite{1937-Whittaker}, Proca \cite{1937-Proca},
  Ruse  \cite{1936-Ruse},
 Taub  \cite{1939-Taub(1), 1939-Taub(2)},  Belinfante   \cite{1939-Belinfante(1),
  1939-Belinfante(2)},
Ivanenko -- Sokolov  \cite{1951-Ivanenko},  Feshbach -- Nikols
\cite{1958-Feshbach(2)}, K\"{a}hler
  \cite{1960-Kahler, 1961-Kahler},
Leutwyler  \cite{1962-Leutwyler},  Klauder \cite{1964-Klauder},
Penney  \cite{1965-Penney}, Cereignani  \cite{1967-Cereignani},
Streater -- Wilde  \cite{1970-Streater}, Pestov  \cite{1971-Pestov,
1978-Pestov, 1983-Pestov}, Osterwalder  \cite{1973-Osterwalder},
 Crumeyrolle  \cite{1975-Crumeyrolle}, Durand \cite{1975-Durand},
 Strazhev et al \cite{1977-Strazhev, 1978-Kruglov, 1978-Strazhev(1),
 1978-Bogush(1),  1978-Bogush(2), 1987-Satikov,  1988-Strazhev(1),
1988-Strazhev(2), 2002-Strazhev, 2002-Tzionenko, 2003-Tzionenko,
2007-Strazhev}, Graf  \cite{1978-Graf}, Benn -- Tucker
\cite{1982-Benn, 1983-Benn(1), 1983-Benn(2), 1983-Benn(3),
1983-Benn, 1985-Benn},   Banks et al   \cite{1982-Banks},
Garbaczewski  \cite{1982-Garbaczewski}, letjuxov  -- Strazhev
\cite{1982-Pletjuxov, 1986-Pletjuxov, 1987-Pletjuxov,
1989-Pletjuxov},
 Holland   \cite{1983-Holland},
Ivanenko et al  \cite{1985-Ivanenko, 1993-Obukhov},
Bullinaria  \cite{1986-Bullinaria},  Blau  \cite{1986-Blau}, Jourjine
\cite{1987-Jourjine},
%Редьков  и др. \cite{1989-Red'kov(2),2000-Red'kov(1), 2005-Red'kov(1), 2005-Red'kov(3)},
Krolikowski \cite{1989-Krolikowski(1),   1989-Krolikowski(2)}, Howe
\cite{1989-Howe},  Nikitin et al  \cite{1995-Nikitin(1)}, Kruglov
\cite{1978-Bogush(2), 2000-Kruglov, 2002-Kruglov, 2004-Kruglov(1)}, Marchuk  \cite{1998-Marchuk,  1999-Marchuk(1),
1999-Marchuk(2), 2001-Marchuk, 2002-Marchuk(1), 2002-Marchuk(2),
2002-Marchuk(3)}, Krivskij et al \cite{2005-Krivskij}.

 Three most interesting points in connection of general covariant extension of the wave equation for this field
 are: in flat Minkowski space there exist tensor and spinor formulations of the theory;
 in the initial tensor form there are presented tensors with different intrinsic parities;
 there exist different views about physical interpretation of the  object: whether it is a composite
 boson or a set of four fermions.
These three point will be of primary importance in the treatment below.

\section{    Spinor and tensor forms of the wave equation}

In Minkowski space-time, the Dirac--K\"{a}hler field is described
by 16-component wave function with transformation properties of 2-rank 4-bispinor $U(x)$ or by equivalent
set of elementary tensor  constituents
$$
U(x)   \qquad   \mbox{or}  \qquad  \{ \Psi (x), \Psi _{i}(x),
\tilde{\Psi }(x),
     \tilde{\Psi }_{i}(x), \Psi _{mn}(x) \} \; ,
$$

\noindent where  $\Psi (x)$ is a scalar;  $\Psi _{i}(x)$ is a vector;
$\tilde{\Psi }(x)$ is a pseudoscalar; $\tilde{\Psi }_{i}(x)$  represents a pseudovector;
  $\Psi _{mn}(x)$ is  an anti-symmetric tensor.  Correspondingly,
  we have two representations for  the wave equation
$$
[i\;  \gamma ^{a} \partial_{a}  - m \; ]\; U(x)  = 0 \; ,
\eqno(2.1)
$$

\noindent and
$$
\partial _{l}\Psi  + m \Psi _{l} = 0 \; , \qquad
\partial _{l} \tilde{\Psi } + m \tilde{\Psi }_{l} = 0 \; ,
\qquad
\partial _{l} \Psi  + \partial_{a} \Psi _{la} - m \Psi _{l} = 0 \; ,
$$
$$
\partial _{l} \tilde{\Psi } -
{1\over 2} \;\epsilon ^{\;\;amn}_{l} \; \partial _{a} \Psi _{mn} -
m \tilde{\Psi }_{l}  = 0  \; ,
$$
$$
\partial _{m} \Psi _{n} - \partial _{n} \Psi _{m} +
\epsilon ^{\;\;\;\;\;ab}_{mn} \;\partial _{a} \tilde{\Psi } _{b} - m
\Psi _{mn} = 0 \; . \eqno(2.2)
$$

Let us detail relation between  2-rank bispinor $U(x)$ and corresponding tensors.
It is well known that any $(4 \times 4)$-matrix can be expanded on  16  Dirac matrices;
and for that expanding it does not  matter  whether the matrix $U$ is a 2-rank bispinor or not.
However, if it is so,  coefficients arising $\{ \Psi , \Psi _{l}, \tilde{\Psi },
\tilde{\Psi }_{l}, \Psi _{mn} \}$   will posses quite definite tensorial properties
with respect to the Lorentz group. Let such a 2-rank bispinor  $U(x)$
is parameterized according to
$$
U(x) = \left [\;  - i \Psi  + \gamma^{l} \; \Psi _{l} +
           i \sigma^{mn}\;  \Psi _{mn} +  \gamma ^{5} \; \tilde{\Psi } +
           i \gamma ^{l} \gamma ^{5} \; \tilde {\Psi }_{l} \; \right ]\; E^{-1} \; ;
\eqno(2.3a)
$$

\noindent
here  $E$ stands for a metrical bispinor matrix with simple properties
$$
E =  \left | \begin{array}{cc}
               \epsilon   &   0  \\ 0   &   \dot{\epsilon}^{-1}
\end{array}  \right | =
     \left | \begin{array}{cc}
               \epsilon_{\alpha \beta}  &   0  \\
                0      &      \epsilon ^{\dot{\alpha}\dot{\beta}}
     \end{array}  \right | =
     \left | \begin{array}{cc}
              i  \sigma^{2}  &   0  \\
                0      &    - i   \sigma^{2}
     \end{array}  \right | \; ,
$$
$$
E^{2} = - I \; , \qquad \tilde{E} = - E \; , \qquad  \mbox{Sp} \;
E = 0 \; ,\qquad
 \tilde{\sigma }^{ab} E = - E \sigma ^{ab} \; .
\eqno(2.3b)
$$
\noindent Inverse to  $(2.3a)$ relations are
$$
\Psi (x) = -{1 \over4i} \; \mbox{Sp} \; [ E U(x) ] \; ,\qquad
\tilde{\Psi }(x) = {1\over 4} \; \mbox{Sp}\;  [E \gamma ^{5} U(x)]
\; ,
$$
$$
\Psi _{l}(x) = {1\over 4}  \; \mbox{Sp} \; [ E \gamma _{l} U(x)]
\; , \qquad \tilde{\Psi }_{l}(x) = {1\over 4i} \; \mbox{Sp}\;  [E
\gamma ^{5}\gamma _{l} U(x)] \; ,
$$
$$
\Psi _{mn}(x)= -{1 \over 2i} \;  \mbox{Sp}\;  [E \sigma _{mn} U(x)
]\; . \eqno(2.3c)
$$

Below we will use also a 2-component spinor formalism;  to this end, it suffices  to choose Dirac matrices in spinor
Weyl basis and specify additionally  notation for constituents of $U(x)$:
$$
U(x) = \left | \begin{array}{cc}
\xi ^{\alpha \beta }(x)  & \Delta ^{\alpha }_{\;\;\dot{\beta}}(x) \\
H_{\dot{\alpha}}^{\;\;\beta}(x) &
\eta_{\dot{\alpha}\dot{\beta}}(x)
        \end{array} \right  |                          \; .
\eqno(2.4a)
$$

\noindent Thus, instead of  $(2.3a)$ we obtain
$$
\Delta (x) = [ \;\Psi _{l}(x) + i \; \tilde{\Psi}_{l}(x) \; ]\;
\sigma^{-l}\dot{\epsilon } \; ,\qquad
H(x) = [ \;\Psi _{l}(x) - i \; \tilde{\Psi }_{l}(x))\;] \;
\bar{\sigma}^{l} \epsilon ^{-1} \; , \eqno(2.4b)
$$
$$
\xi (x) = [ \; - i \;  \Psi (x) - \tilde{\Psi} (x) + i \; \Sigma
^{mn} \; \Psi _{mn}(x)\;  ] \; \epsilon ^{-1} \; ,
$$
$$
\eta (x) = [\; - i \; \Psi (x) + \tilde{\Psi} (x) +
 i\;  \bar{\Sigma}^{mn}\; \Psi _{mn}(x)\; ]\; \dot{\epsilon } \;  ,
$$

\noindent and inverse relations
$$
\Psi _{l}(x) + i \; \tilde{\Psi} _{l}(x)  = {1\over 2} \;
\mbox{Sp}\;  [\; \dot{\epsilon }^{-1} \; \sigma _{l} \;\Delta(x)
\; ] \; , \qquad \Psi _{l}(x) - i \; \tilde{\Psi }_{l}(x)  =
 {1\over 2} \; \mbox{Sp} \;  [\;  \epsilon \bar{\sigma}_{l} \; H(x)\; ] \; ,
$$
$$
-i \; \Psi(x)  - \tilde{\Psi}(x)   = {1\over 2} \; \mbox{Sp}\;
[\; \epsilon  \; \xi(x)\;  ] \; , \qquad -i \; \Psi(x)  +
\tilde{\Psi }(x)  = {1\over 2} \; \mbox{Sp} \; [ \; \dot{\epsilon
}^{-1} \; \xi(x) \; ] \; ,
$$
$$
  - i \; \Psi ^{kl}(x) + {1\over 2}\; \epsilon ^{klmn} \;\Psi _{mn}(x) =
\mbox{Sp}\; [ \; \epsilon \;  \Sigma ^{kl} \xi(x)\; ] \; ,
$$
$$
 - i\;  \Psi ^{kl}(x) - {1\over 2}\; \epsilon ^{klmn}\; \Psi _{mn}(x) =
\mbox{Sp} \; [ \; \dot{\epsilon }^{-1} \; \bar{\Sigma}^{kl} \;
\xi(x) \; ] \; . \eqno(2.4c)
$$

Dirac--K\"{a}hler equation in 2-spinor form
looks as follows
$$
i\sigma ^{a} \;\partial_{a}\; \xi (x) = m \;   H(x) \; , \qquad
i\bar{\sigma}^{a} \; \partial_{a} \;  H(x) = m \; \xi (x) \; ,
$$
$$
i \bar{\sigma}^{a} \; \partial_{a} \; \eta (x) = m \; \Delta (x)
\; , \qquad i \sigma ^{a} \; \partial_{a} \; \Delta (x)    = m \;
\eta (x) \; . \eqno(2.5)
$$

Now let us consider a general covariant form.
First, we turn to the 4-spinor approach -- according to the well known recipe by Tetrode--Weyl--Fock--Ivanenko
eq. (2.1)  should be  changed into
$$
\left [ \; i \gamma ^{\alpha }(x)\; ( \partial_{\alpha} \; + \;
B_{\alpha }(x) )\; - \; m\; \right ] \; U(x) = 0 \; ; \eqno(2.6)
$$

\noindent connection  $B_{\alpha }(x) $ is defined by
$$
B_{\alpha }(x) = {1\over 2} J^{ab} e^{\beta }_{(a)}(x)
 \nabla _{\alpha }e_{(b)\beta }(x) =  \Gamma _{\alpha }(x) \otimes  I +
I \otimes  \Gamma_{\alpha}(x) \; ,
$$

\noindent where $J^{ab} = [\sigma ^{ab} \otimes I + I \otimes \sigma
^{ab} ]$ stand for generators for bispinor representation of the Lorentz group.
From   (2.6)  it follow 2-spinor form of equations for Dirac--K\"{a}hler field
$$
i \sigma ^{\alpha }(x)\; [ \; \partial_{\alpha} \; + \;
 \Sigma _{\alpha }(x) \otimes I + I \otimes  \Sigma _{\alpha }(x)\; ] \;
 \zeta (x) = m\; H(x) \; ,
$$
$$
i \bar{\sigma}^{\alpha }(x)\; [\; \partial_{\alpha} \; + \;
 \bar{\Sigma}_{\alpha }(x) \otimes I +
 I \otimes  \Sigma _{\alpha }(x)\; ] \; H(x) = m \; \xi (x) \; ,
$$
$$
i \bar{\sigma}^{\alpha }(x)\; [\; \partial_{\alpha} \; + \;
\bar{\Sigma}_{\alpha }(x) \otimes  I + I \otimes
\bar{\Sigma}_{\alpha }(x)\; ] \; \eta (x) = m \; \Delta (x) \; ,
$$
$$
i \sigma ^{\alpha }(x)\; [\; \partial_{\alpha} \; + \; \Sigma
_{\alpha }(x) \otimes  I + I \otimes \bar{\Sigma}_{\alpha }(x)\;]
\; \Delta (x) = m \; \eta (x) \; . \eqno(2.7)
$$

Eqs. (2.6)  and  (2.7)  posse symmetry  with respect to local Lorentz group:
if $U(x)$   is subject to local Lorentz transformation
$$
U'(x) = [\; S(k(x),k^{*}(x)) \otimes  S(k(x),k^{*}(x)) \; ] \;
U(x) \; , \eqno(2.8a)
$$

\noindent then the new field function  $U'(x)$, or set of new 2-spinors
$[\;\xi '(x),  \; \eta'(x),\;  \Delta '(x), \; H'(x)\;]$,  will obey a wave equation
of the same type as before
$$
[ \;i \gamma'^{\alpha }(x)\; ( \partial_{\alpha}\; + \;
B'_{\alpha}(x) )\;  -\; m\; ]\;  U'(x) = 0 \; , \eqno(2.8b)
$$

\noindent where  primed  $\gamma '^{\alpha }(x)$   and  $B'_{\alpha}(x)$
are constructed with the help of primed tetrad
$e'^{\alpha }_{b)}(x)$, related to the initial one by Local Lorentz transformation
$$
e'^{\alpha }_{(b)}(x)  = L_{b}^{\;\;a}(k(x),k^{*}(x)) \;
\;e^{\alpha }_{(a)}(x) \; .
$$

\noindent
 This symmetry prove correctness of the equation under consideration: the symmetry describes a gauge freedom
 in choosing an explicit form of  the tetrad.

In addition, there exists   discrete  symmetry. Indeed,
if  $U(x)$ is subject to the following discrete operation
$$
U'(x) = [  i \gamma^{0} \otimes   i \gamma ^{0} ] \; U(x)\; \qquad
\mbox{or} \qquad \left | \begin{array}{cc}
\xi '(x) &   \Delta'(x) \\
H'(x)   &   \eta ' (x)
\end{array} \right |
 =
\left | \begin{array}{cc}
 - \eta  (x)  & -H(x) \\
 -\Delta (x)  & -\xi(x)
 \end{array} \right | \; ,
\eqno(2.9a)
$$

\noindent then the new  wave function  $U'(x)$ (new set of 2-spinors)
will obey an equation of the same form  (2.6) (or (2.7)),   but now  constructed on the base of
a tetrad  $e'^{\alpha
}_{(b)}(x)$, Р-reflected to the initial
$$
e'^{\alpha }_{(b)}(x)  = L^{\;(p) a}_{b} \; e^{\alpha}_{a}(x) \; ,
\qquad L^{(p) a}_{b} = \mbox{diag} \; (+1, -1, -1, -1) \; .
\eqno(2.9b)
$$

 With respect to general coordinate transformations,
 the wave function  $U(x)$ behaves as a scalar (similarly as a wave function  $\Psi (x)$ in the Dirac equation does)
Correspondingly, the term   $\partial_{\alpha }  U(x)$ represents a general covariant vector and eq.
(2.6)  is correct  in the sense of general covariance.

Now we turn to extending the tensor form of equations
(2.2).  We face here a rather specific problem. Indeed, a formal change
$$
\partial_{l}   \Longrightarrow   \nabla _{\alpha } \; , \qquad
\Psi _{i}(x) \; \Longrightarrow \; \Psi _{\alpha }(x)\; , \;
$$
$$
\tilde{\Psi} _{i}(x) \; \Longrightarrow \; \tilde{\Psi} _{\alpha
}(x)\; , \qquad \Psi _{ij}(x)\; \Longrightarrow \; \Psi
_{\alpha\beta}(x)
$$

\noindent leads to appearing some vagueness:  it is not clear how we should  distinguish
between two functions
$\Psi _{\alpha }(x)$  and $\tilde{\Psi }_{\alpha }(x)$ -- because they have one the same  index  $\alpha $,
the sign of covariant
vector. Nevertheless, making such a formal generalization we get the system
$$
\nabla ^{\alpha } \Psi _{\alpha }(x) + m \Psi (x) = 0 \; , \qquad
\nabla^{\alpha} \tilde{\Psi}_{l}(x) + m \tilde{\Psi}(x) = 0 \; ,
$$
$$
\nabla _{\alpha }\Psi (x) + \nabla ^{\beta } \Psi _{\alpha \beta
}(x)  - m \Psi _{\alpha }(x) = 0 \; , \;
$$
$$
\nabla_{\alpha } \tilde{\Psi}(x)  - {1\over 2} \epsilon
^{\;\;\beta \rho \sigma }_{\alpha }(x) \; \nabla _{\beta } \Psi
_{\rho \sigma }(x) - m \tilde{\Psi }_{\alpha }(x)  = 0 \; ,
$$
$$
 \nabla _{\alpha } \Psi _{\beta }(x) - \nabla _{\beta } \Psi _{\alpha }(x) +
\epsilon ^{\;\;\;\;\rho \sigma }_{\alpha \beta }(x)\;
\nabla_{\rho} \tilde{\Psi }_{\sigma }(x)  - m \Psi _{\alpha \beta
}(x) = 0 \; . \eqno(2.10)
$$

Resolving the problem of distinguishing between
 $\Psi _{\alpha }(x)$  and   $\tilde{\Psi }_{\alpha }(x)$, also $\Psi (x)$  and $\tilde{\Psi }(x)$,
also  determining a covariant Levi-Civita  object, can be found for comparing
eq.  (2.10) with  eq. (2.6).

We will  demonstrate that from
eq.  (2.6) it follows eqs.    (2.10), if instead of  $U(x)$   in  (2.6)
we substitute expansion of the matrix $U(x)$  in terms of tetrad tensor constituents
and then translate equations to covariant tensors according to
$$
\Psi _{\alpha }(x) = e^{(i)}_{\alpha }(x) \; \Psi _{i}(x)\; ,
\qquad \tilde{\Psi }_{\alpha }(x) = e^{(i)}_{\alpha }(x)\;
\tilde{\Psi }_{i}(x) \; ,\;
$$
$$
\Psi _{\alpha \beta }(x) = e^{(m)}_{\alpha }(x) e^{(n)}_{\beta
}(x)\; \Psi _{mn}(x) \; , \eqno(2.11a)
$$

\noindent and a covariant Levi--Civita object is defined as follows
$$
\epsilon ^{\alpha \beta \rho \sigma }(x) = \epsilon^{abcd} \;
e^{\alpha }_{(a)}(x) \; e^{\beta }_{(b)}(x)\; e^{\rho }_{(c)}(x)\;
 e^{\sigma }_{(d)}(x) \; .
\eqno(2.11b)
$$

\noindent At this we note that the relevant similar functions entering eqs.
 (2.10) differ in their transformation properties with respect to tetrad $P$-reflection:
 $\Psi (x),\; \Psi _{\alpha
}(x), \; \Psi _{\alpha \beta }(x)$ are tetrad scalars;
$\tilde{\Psi }(x),\; \tilde{\Psi }_{\alpha }(x)$  are tetrad pseudoscalars.

Let us explain calculations proving this.
First, eq.  (2.6) is written in the form (the symbol $\sim$ designates matrix transposition)
$$
[\; i \gamma ^{\alpha } \; \partial_{\alpha} \; U + i
\gamma^{\alpha } \;  \Gamma _{\alpha }(x)\;  U  + i \gamma
^{\alpha} \; U \; \tilde{\Gamma}_{\alpha } - m \; U\; ] = 0 \; ,
\eqno(2.12a)
$$

\noindent then eq. $(2.12a)$ is translated to
$$
[\; i \gamma ^{c} \; e^{\alpha }_{(c)} \; \partial_{\alpha}\; U +
{i \over 2} \; \gamma _{abc} \; \gamma ^{c} \; \sigma ^{ab} \; U +
{i \over 2}\;  \gamma _{abc} \; \gamma ^{c}\; U\; \tilde{\sigma
}^{ab} - m \;U\;  ]
 = 0 \; .
\eqno(2.12b)
$$

\noindent Further, into $(2.12b)$  we substitute expansion for $U$ in term of local tetrad tensors
$$
\left \{ \; i \gamma ^{c} \;  e^{\beta }_{(c)} \; \partial_{\beta}
\; [\; - i \Psi  +  \gamma ^{l} \; \Psi _{l} + i \sigma ^{mn} \;
\Psi _{mn} + \gamma ^{5} \tilde{\Psi} + i \gamma ^{l}\gamma ^{5}
\tilde{\Psi }_{l}]\; E^{-1} \;  \right.
$$
$$
+ { i \over 2} \gamma _{abc} \gamma ^{c} \sigma ^{ab} \; [- i \Psi
+ \gamma ^{l} \Psi _{l} + i \sigma ^{mn} \Psi _{mn} + \gamma ^{5}
\tilde{\Psi} + i \gamma ^{l} \gamma ^{5} \tilde{\Psi }_{l}\; ] \;
E^{-1}
$$
$$
+ {i \over 2} \gamma _{abc} \gamma ^{c}\; [\; - i \Psi  + \gamma
^{l} \Psi _{l} +
 i \sigma ^{mn} \Psi _{mn} + \gamma ^{5} \tilde{\Psi} +
 i \gamma ^{l}\gamma ^{5} \tilde{\Psi }_{l} \; ]\;  E^{-1} \tilde{\sigma }^{ab}
$$
$$
 \left. -  m \; [\; - i \Psi  + \gamma ^{l} \Psi _{l} + i \sigma ^{mn} \Psi _{mn} +
 \gamma ^{5} \tilde{\Psi} + i \gamma ^{l} \gamma ^{5} \tilde{\Psi }_{l}\; ]
 E^{-1}\; \right \}  = 0\; .
\eqno(2.12c)
$$

\noindent Now, acting subsequently from the left by operators
$$
\mbox{Sp}\; (E\times \; , \qquad \mbox{Sp}\; (E \gamma ^{5} \times
\; , \qquad \mbox{Sp}\; (E \gamma ^{k}\times \; , \qquad
\mbox{Sp}\; (E \gamma ^{5}\gamma ^{k} \times \; , \qquad
\mbox{Sp}\; (E \sigma ^{kd} \times
$$

\noindent and using known formulas for traces of relevant combinations of Dirac matrices
we arrive at
$$
e^{(l)\alpha} \; \partial_{\alpha} \psi _{l} + \gamma
^{c}_{\;\;lc} \; \Psi ^{l} +  m \Psi = 0 \; ,
$$
$$
 e^{(l)\alpha } \; \partial_{\alpha}
 \tilde{\Psi }_{l}  + \gamma ^{c}_{\;\;lc} \; \tilde{\Psi }^{l}  +
m \tilde{\Psi}  = 0 \; ,
$$
$$
 e^{(k)\alpha } \; \partial_{\alpha} \Psi  +
 e^{\alpha }_{(c)} \; \partial_{\alpha} \Psi^{kc} +
\gamma ^{k}_{\;\;mn} \; \Psi ^{mn} +
 \gamma ^{c}_{\;\;lc} \; \Psi ^{kl}  - m \Psi ^{k} = 0 \; ,
$$
$$
e^{(k)\alpha } \; \partial_{\alpha} \tilde{\Psi} - {1 \over 2}
\epsilon ^{kcmn} \; e^{\alpha }_{(c)} \;
\partial_{\alpha} \Psi _{mn} +
\epsilon ^{kcmn} \;  \gamma ^{n}_{\;\;bc} \; \Psi _{mn} - m
\tilde{\Psi }^{k}
 = 0  \; ,
$$
$$
 e^{(d)\alpha} \;  \partial_{\alpha} \Psi ^{k} -
e^{(k)\alpha} \; \partial_{\alpha} \Psi ^{d}  +
 (\gamma ^{\;\;dk}_{l}  - \gamma ^{\;\;kd}_{l}) \Psi ^{l}
$$
$$
+ \epsilon ^{dkcl} \; e^{\alpha }_{(c)} \; \partial_{\alpha}
 \tilde{\Psi }_{l} + \epsilon ^{acdk} \; \gamma _{bac}
\tilde{\Psi }^{b} - m \Psi ^{dk}  = 0 \; ; \eqno(2.12d)
$$

\noindent they represent written in tetrad components $(2.11a,b)$ eqs.  (2.10).

One important point should be specially emphasized: during calculation, a Levi--Civita object  $\epsilon ^{abcd}$
arose in $(2.12d)$  as a direct result of the use of a trace formula for product of three Dirac matrices,
  so this quantity $\epsilon ^{abcd}$  is not a tensor with respect to the lorentz group,
  it is rather just a fixed 4-index object.

It is readily to demonstrate that the combination
$$
\epsilon ^{\alpha \beta \rho \sigma }(x) =  \; \epsilon ^{abcd} \;
e^{\alpha }_{(a)}(x) \; e^{\beta }_{(b)}(x)  \;
 e^{\rho }_{(c)}(x) \; e^{\sigma }_{(d)}(x)
\eqno(2.13a)
$$

\noindent represents a tetrad pseudoscalar. Indeed, let us compare
 $\epsilon ^{\alpha \beta \rho \sigma }(x)$   and
 $\epsilon'^{\alpha \beta \rho \sigma }(x)$, constructed on the base of tetrads
 $e^{\alpha }_{(a)}(x)$   and   $e'^{\alpha }_{(a)}(x)$ respectively. We have

$$
\epsilon'^{\alpha \beta \rho \sigma }(x) = \epsilon ^{abcd} \;
e'^{\alpha}_{(a)}(x) \;
                    e'^{\beta }_{(b)}(x) \;
                    e'^{\rho }_{(c)}(x)  \;
                    e'^{\sigma}_{(d)}(x) \; ,
$$

\noindent    or

$$
\epsilon'^{\alpha \beta \rho \sigma }(x) = \epsilon ^{abcd} \;
L^{\;\;i}_{a}(x)\; L^{\;\;j}_{b}(x)\; L^{\;\;m}_{c}(x) \;
L^{\;\;n}_{d}(x) \;
                    e^{\alpha }_{(i)}(x)   \;
                    e^{\beta }_{(j)} (x)   \;
                    e^{\rho }_{(m)}  (x)   \;
                    e^{\sigma }_{(n)}(x) =
$$

$$
\left [ \; \epsilon ^{abcd} \; L_{ai}(x) \; L_{bj}(x) \; L_{cm}(x)
\; L_{dn}(x)\; \right ] \;
   e^{(i)\alpha}(x) \; e^{(j)\beta }(x) \; e^{(m)\rho }(x) \; e^{(n)\sigma }(x)\; .
$$

\noindent With the use of the known identity  \cite{1973-Landau}
$$
(\epsilon ^{abcd}\; A_{ai} \; A_{bj}\;  A_{cm} \; A_{dn} ) = -
 \det [A_{ab}] \times  \epsilon _{ijmn} \;,
 $$

 \noindent
 we get a transformation low for
$\epsilon ^{\alpha \beta \rho \sigma }(x)$  with respect to local  tetrad transformations:
$$
\epsilon'^{\alpha \beta \rho \sigma }(x)
   = - \det [L_{ai}(x)] \; \epsilon ^{\alpha \beta \rho
   \sigma }(x) \; .
\eqno(2.13b)
$$

\noindent From $(2.13b)$ it follows that under tetrad $P$-reflection covariant Levi-Civita object
$(2.13a)$ behaves as a tetrad pseudoscalar
$$
\epsilon ^{(p)\alpha \beta \rho \sigma}(x) = (-1) \; \epsilon
^{\alpha \beta \rho \sigma }(x) \; . \eqno(2.13c)
$$

\noindent
One can notice that in each equation in   (2.10),    there are combined terms with
equal transformation propertied with respect to the tetrad $P$-reflection.

The system (2.10) can be translated to the form in which
all the component of the wave function are tetrad scalars:
$$
\Phi (x) = \{ \; \Psi (x), \; \Psi _{\alpha }(x)  ,\;  \Psi
_{\alpha \beta }(x) ,\;
              \Psi _{\alpha \beta \rho }(x)
              $$
              $$
              =
    \epsilon _{\alpha \beta \rho \sigma }(x) \; \tilde{\Psi }^{\sigma }(x)  ,\;
    \Psi _{\alpha \beta \rho \sigma }(x) =
\epsilon _{\alpha \beta \rho \sigma }(x)\; \tilde{\Psi }(x) \; \}
\; .
\eqno(2.14a)
$$

\noindent then the Dirac--K\"{a}hler equation reads
$$
\nabla ^{\rho} \Psi _{\rho} - m \Psi   = 0 \; ,
$$
$$
\nabla ^{\rho } \Psi _{\rho \alpha } + \nabla _{\alpha } \Psi
 + m \Psi _{\alpha } = 0  \; ,
$$
$$
\nabla ^{\rho } \Psi _{\rho \alpha \beta }  + \nabla _{\alpha }
\Psi _{\beta }  - \nabla _{\beta } \Psi _{\alpha } - m \Psi
_{\alpha \beta }  = 0 \; ,
$$
$$
\nabla ^{\rho }\Psi _{\rho \alpha \beta \sigma } + \nabla _{\alpha
}\Psi _{\beta \sigma } - \nabla _{\beta }\Psi _{\alpha \sigma } -
\nabla _{\sigma }\Psi _{\beta \alpha } +
 m \Psi _{\alpha \beta \sigma } (x)  = 0  \; ,
$$
$$
\nabla _{\rho }\Psi _{\alpha \beta \sigma }  - \nabla _{\alpha
}\Psi _{\rho \beta \alpha }  - \nabla _{\beta }\Psi _{\alpha \rho
\sigma }  - \nabla _{\sigma }\Psi _{\alpha \beta \rho }  - m \Psi
_{\rho \alpha \beta \sigma }  = 0 \; . \eqno(2.14b)
$$

Deriving  $(2.14b)$ from  (2.10), one should take into account that
covariant derivative of the covariant Levi--Civita tensor   vanishes identically
$$
\nabla _{\mu } \; \epsilon ^{\alpha \beta \rho \sigma }(x) = 0 \;
. \eqno(2.15)
$$

\noindent Let us prove it. By symmetry reason, it suffices to prove only one relation
 $\nabla _{\mu }   \epsilon _{0123}(x) = 0$. In accordance with definition
 we have

$$
\nabla _{\mu } \; \epsilon _{0123}(x) = [ \; \partial_{\mu}  \;
 \epsilon _{0123}(x) \; - \;  ( \Gamma ^{\nu }_{\mu 0} \; \epsilon _{\nu 123}(x) \; + \;
 \Gamma ^{\nu }_{\mu 1} \; \epsilon _{0\nu 23}(x)
$$

$$
 +\Gamma ^{\nu }_{\mu 2} \; \epsilon _{01\nu 3}(x) \; + \;
 \Gamma ^{\nu }_{\mu 3} \; \epsilon _{012\nu }(x)) \; ] =
\partial_{\mu } \; \epsilon _{0123}(x) \; - \;
\Gamma ^{\alpha }_{\mu \alpha } \; \epsilon _{0123}(x) \; .
$$

\noindent Let us  specify  the first term  $ \partial_{\mu} \; \epsilon _{0123}(x)$, where
$$
\epsilon _{0123}(x) = - \;\epsilon _{0123} \;
\mbox{det}\; [e_{(a)\alpha} (x)] \;;
$$
 with the use of the  known identity \cite{1973-Landau}
$$
\partial_{\mu}\; A = A \; ( A^{-1}_{ji}  \partial_{\mu} \; A_{ij}) \; , \qquad
A = \mbox{det} \;[A_{ij}]
$$

\noindent  and allowing for that  the inverse to $e_{(a)\alpha }$ is a matrix $e^{\beta (b)}$,we get
$$
e(x) = \det [e_{(a)\alpha}(x)] \; , \qquad  \partial_{\mu} \; e(x)
=
  e(x) \;  e^{\alpha(a)}(x) \; \partial_{\mu} \; e_{(a)\alpha}(x) \; .
$$

\noindent Therefore,
$$
\partial_{\mu} \; \epsilon _{0123}(x) =
\epsilon _{0123}(x) \; [\; e^{\alpha (a)}(x) \partial_{\mu} \;
e_{(a)\alpha}(x) \; ] \; .
$$

\noindent In turn, for   $\Gamma ^{\alpha }_{\mu \alpha}(x)$  we have

$$
\Gamma ^{\alpha }_{\mu \alpha}(x)  = {1 \over 2}
 g^{\alpha \rho }(x) \; \Gamma _{\rho ,\mu \alpha }(x) =
{1 \over 2} g^{\alpha \rho }x) \; [ \;
       \partial_{\mu}\; g_{\rho \alpha}(x)\; + \;
       \partial_{\alpha} g_{\rho \mu }(x) \; + \;
       \partial_{\rho} \; g_{\mu \alpha}(x) \; ]
$$
$$
={1\over 2} g^{\alpha \beta }\; [\;
\partial_{\mu} \; (\; e_{(i)\rho}(x)  e^{(i)}_{\alpha}(x)\; ) \; + \;
\partial_{\alpha} \; (\; e_{(i)\rho}(x) e^{(i)}_{\mu}(x)\; ) \; + \;
\partial_{\rho} \;  (\; e_{(i)\mu}(x) \; e^{(i)}_{\alpha}(x)\; ) \; ] \; ,
$$

\noindent from whence after simple calculation we derive
$$
\Gamma ^{\alpha }_{\mu \alpha}(x) = e^{\alpha }_{(i)}(x) \;
\partial_{\mu } \; e^{(i)}_{\alpha}(x) \;.
$$

\noindent Thus, we prove the needed identity
$$
\nabla _{\mu } \; \epsilon ^{\alpha \beta \rho \sigma }(x) = 0 \;
.
$$

\section{ On two different covariant Levi-Civita objects}

Let us recall a standard view on covariant Levi-Civita object -- it is defined
 \cite{1973-Landau} as follows
$$
E_{\alpha \beta \rho  \sigma}(x) \equiv  + \sqrt{-g(x)} \; \;
\epsilon_{\alpha \beta \rho  \sigma } \;  ,  \qquad E^{\alpha
\beta \rho  \sigma}(x) \equiv {1 \over  +\sqrt{-g(x)}} \;\;
\epsilon^{\alpha \beta \rho  \sigma } \;  ,
$$
$$
E^{\alpha \beta \rho  \sigma}(x) = g^{\alpha \mu} (x) g^{\beta
\nu} (x) g^{\rho \gamma}(x) g^{\sigma \delta}(x)\; E_{\mu \nu
\gamma \delta} (x)\; , \eqno(3.1a)
$$

\noindent where   $g(x)$ is a determinant of a metric tensor
$g_{\alpha \beta}(x)$; and
 $E_{0123} (x) = + \sqrt{-g(x)} $.
This definition does not depend on tetrads at all, which means that
$ E_{\alpha \beta \rho  \sigma}(x) $ is a tetrad scalar.
To have the covariant Levi-Civita object invariant with respect to arbitrary coordinate chnges
we must assume  that  the  object $E_{\alpha \beta \rho
\sigma}(x) $  transform as a pseudotensor, that is we add in relevant transformation low
an additional a special factor
$\mbox{sgn} \; \Delta (x) $
$$
\mbox{sgn} \; \Delta (x) = {  \Delta (x) \over   \mid \Delta (x)
\mid } \; , \qquad \Delta(x)  \equiv  \mbox{det} \; [{\partial
x'^{\alpha} \over
\partial x^{\alpha}}]   \; ,
$$
$$
E_{\alpha' \beta' \rho' \sigma'} = {  \Delta (x) \over   \mid
\Delta (x) \mid }  \; {\partial x^{\alpha}\over \partial
x^{\alpha'}}\; {\partial x^{\beta}\over \partial x^{\beta'}}\;
{\partial x^{\rho}\over \partial x^{\rho'}}\; {\partial
x^{\sigma}\over \partial x^{\sigma'}}\; E_{\alpha \beta \rho
\sigma} \; . \eqno(3.1b)
$$

Above, in the frames of the tetrad formalism, the quantity $ \epsilon_{\alpha
\beta \rho  \sigma } $  was introduced by  $(2.11b)$; so it is an ordinary covariant tensor with 4 indices and
in the same time it is a tetrad pseudoscalar. Two objects,  $
\epsilon_{\alpha \beta \rho  \sigma}(x)$ and $E_{\alpha \beta \rho
\sigma}$ were defined independently from each other, therefore they may not coincide.
However, quite definite relation between them exists, let us detail this point.

First of all, let us transform the tetrad based Levi-Civita tensor  $ \epsilon_{\alpha \beta \rho  \sigma }(x) $
to a different form similar to  $(3.1a)$:
$$
\epsilon_{\alpha \beta \rho  \sigma } (x) = \epsilon ^{abcd} \;
e_{(a) \alpha }(x) \; e_{(b) \beta }(x) \; e_{(c) \rho }(x) \;
e_{(d)\sigma }(x) =
  -\; e(x) \; \epsilon_{\alpha \beta \rho  \sigma } \; ,
  $$
  $$
 e(x) \equiv \mbox{det}  \; [ \; e_{(a)\alpha}(x) \; .
\eqno(3.2)
$$

\noindent   For instance, in the case of flat Minkowski space,
using a diagonal tetrad   $e_{(a)}^{\alpha }(x) =
\delta _{a}^{\alpha }$,  we get $e(x) = -1$, and further derive
$\epsilon_{\alpha \beta \rho  \sigma } (x) = + \epsilon_{\alpha
\beta \rho  \sigma }\;$.

It is easy to obtain relation relating determinants of the tetrad and metric tensor
$$
e_{(a)\alpha}(x) \; e_{(b)\beta}(x) \; g^{ab} = g_{\alpha
\beta}(x) \; \Longrightarrow \; -e^{2}(x) = g(x) \;,
$$

\noindent from whence it follows
$$
e(x) = + \sqrt{-g(x)}  \qquad  \mbox{and} \qquad e(x) = -
\sqrt{-g(x)} \; .
$$

Taking solution as  $e(x) = - \sqrt{-g(x)}$, we arrive at the tetrad based definition for Levi-Civita tensor (2.3);
so it is equivalent to definition according to
 $(3.1a)$. However, a tetrad determinant  can be positive as well, in this case
 two definition are not equivalent
  -- they differ in sign.

Note a useful formula
$$
\epsilon^{\mu \nu \gamma \delta}(x) = - e(x) \; [\; g^{\mu
\alpha}(x) g^{\nu \beta}(x) g^{\gamma \rho}(x) g^{\delta
\sigma}(x) \; \epsilon_{\alpha \beta \rho  \sigma }\; ]
$$
$$
= + e(x) \; \mbox{det}\; [g^{\alpha \beta}(x)] \; \epsilon^{\mu
\nu \gamma \delta} = +e(x) \; {1 \over g(x) } \; \epsilon^{\mu \nu
\gamma \delta}  = -{ 1 \over e(x)}\; \epsilon^{\mu \nu \gamma
\delta} \; . \eqno(3.3)
$$

Let us specify transformation properties for  $e(x) =
\mbox{det} \; [ e_{(a)\alpha}(x)] $.  Under general coordinate transformations
it behaves
$$
x^{\alpha} \; \Longrightarrow \; x'^{\alpha} \; , \;\; e'(x') =
\mbox{det} \; [\; {\partial  x^{\alpha} \over \partial
x'^{\alpha}} \; e_{(a)\alpha}(x)\; ] = {1 \over \Delta (x) } \;
e(x) \; ; \eqno(3.4a)
$$

\noindent with respect to tetrad changes it is a pseudoscalar
$$
e_{(a)\alpha}(x)  \; \Longrightarrow \; e'_{(a)\alpha}(x) \; ,
\;\; e'(x) = \mbox{det} \; [ L_{a}^{\;\;b} e_{(b)\alpha}(x)] =
\mbox{det}\; [ L_{a}^{\;\;b} ]\; e(x) \; . \eqno(3.4b)
$$

\noindent Let us introduce special quantity
$$
J (e) = - \; {\mbox{det}\; [e_{(a)\alpha}(x)] \over \mid
\mbox{det}\;    [e_{(a)\alpha}(x)] \mid } = - \; {e(x) \over \mid
e(x) \mid } \; . \eqno(3.5a)
$$

\noindent which transforms as follows
$$
x^{\alpha} \; \Longrightarrow \; x'^{\alpha} \; , \;\; J [ e'(x')
]  = {\Delta (x) \over \mid \Delta (x) \mid } \; J [e(x)] \; ,
\eqno(3.5b)
$$
$$
e_{(a)\alpha}(x)  \; \Longrightarrow \; e'_{(a)\alpha}(x) \; ,
\;\; J [ e'(x) ]  =  det (L_{a}^{\;\;b}) \; J [e(x)] \; ;
\eqno(3.5c)
$$

\noindent   $J [e(x)]$ is a tetrad pseudoscalar, and a coordinate pseudoscalar

Collecting  results together
$$
\epsilon_{\alpha \beta\rho \sigma}(x) = - e\; \epsilon_{\alpha
\beta\rho \sigma}\;,
$$
$$
E_{\alpha \beta\rho \sigma}(x) =  \sqrt{- g} \; \epsilon_{\alpha
\beta\rho \sigma}\;,
$$
$$
e^{2} = -g \; , \qquad +\sqrt{-g} = { e \over \mid e \mid } \; e
\; ,
$$

\noindent  we readily find relation between two Levi-Civita tensors
$$
E_{\alpha \beta\rho \sigma}(x) = { e \over \mid e \mid } \; e
\;\epsilon_{\alpha \beta\rho \sigma}= - { e \over \mid e \mid } \;
\epsilon_{\alpha \beta\rho \sigma}(x)= J [e(x)] \;
\epsilon_{\alpha \beta \rho  \sigma } (x)  \; .
$$

Let us turn back to $(2.11a)$. Instead of   $\tilde{\Psi}(x),
\tilde{\Psi}_{\alpha}(x)$  one can introduce new variables
$$
\bar{\Psi}(x) = J(e) \; \tilde{\Psi}(x)\; , \qquad
\bar{\Psi}_{\alpha}(x) = J(e) \; e_{\alpha}^{(a)}(x)
\tilde{\Psi}_{a}(x) \; , \eqno(3.6a)
$$

\noindent and instead of   $\epsilon_{\alpha \beta \rho  \sigma } (x)$
(2.11b) one may determine another quantity
$$
E_{\alpha \beta \rho  \sigma } (x) = J [e(x)] \; \epsilon_{\alpha
\beta \rho  \sigma } (x)  \; . \eqno(3.6b)
$$

\noindent Correspondingly, the main system   of tensor equations can be presented as
follows  (compare with  (2.10))
$$
\nabla ^{\alpha } \Psi _{\alpha } + m\; \Psi  = 0 \;  ,
$$
$$
\nabla^{\alpha} \bar{\Psi}_{l} + m \;\bar{\Psi} = 0 \; ,
$$
$$
\nabla _{\alpha }\Psi  + \nabla ^{\beta } \Psi _{\alpha \beta }  -
m \;\Psi _{\alpha } = 0 \; ,
$$
$$
\nabla_{\alpha } \bar{\Psi}  - {1\over 2} E ^{\;\;\beta \rho
\sigma }_{\alpha }(x) \; \nabla _{\beta } \Psi _{\rho \sigma } - m
\;\bar{\Psi }_{\alpha }  = 0 \; ,
$$
$$
 \nabla _{\alpha } \Psi _{\beta } - \nabla _{\beta } \Psi _{\alpha } +
E^{\;\;\;\;\rho \sigma }_{\alpha \beta }(x)\; \nabla_{\rho}
\bar{\Psi }_{\sigma }  - m \;\Psi _{\alpha \beta } = 0 \; .
\eqno(3.7)
$$

\noindent Here  $\Psi (x), \Psi_{\alpha}(x),\Psi_{\alpha
\beta}(x)$ are general covariant tensors, whereas $\bar{\Psi} (x),
\bar{\Psi}_{\alpha}(x), E^{\rho \sigma \alpha \beta }(x)$  are general covariant pseudotensors;
all six objects are tetrad scalars.

\begin{quotation}

{
Thus, when describing tensor components for Dirac--K\"{a}hler field one can  use alternatively
 both methods.
Evidently, classification of the components through their tetrad properties
is more preferable  because it has clear Lorentzian status (as spin and mass).

It should be noted additionally that classification for tensor quantities in the frames of the full
Lorentz group within Minkowski space-time assumes four different possibilities distinguished by adding special factors
 in transformation low
$$
1\; , \qquad \mbox{det} \; (L^{a}_{\;\;b} ), \qquad \mbox{sgn} \;
(L^{0}_{\;\;0} )\; , \qquad \mbox{sgn} \; (L^{0}_{\;\;0}) \;
\mbox{det} \; (L^{a}_{\;\;b} )\; .
$$

\noindent It is not clear how that Lorentz group based classification
can be described  in terms of a pure  general covariant  theory without tetrad formalism.

}

\end{quotation}

\section{ On fermion interpretation for Dirac--K\"{a}hler field
}

The Dirac--K\"{a}hler equation in arbitrary curved space-time
$$
[\; i \gamma ^{\alpha }(x)\; (\; \partial _{\alpha} \; + \;
B_{\alpha }(x)\; )\;  - \; m \; ]\; U (x) = 0 \; . \eqno(4.1a)
$$

\noindent does not split up into four  independent equations for particles with spin 1/2:
$$
[ \; i \gamma ^{\beta }(x)\;  (\partial_{\beta} \; + \; \Gamma
_{\beta }(x))\; - \;  m \; ]\; \Psi ^{(i)}(x) = 0 \; , \qquad (i =
1,\; 2,\; 3,\; 4)\; ; \eqno(4.1b)
$$

In other words, these two models  are completely different  in any curved space-time model.

Let us consider eqs. $(4.1b)$ in more detail.  Relevant four local  bispinor fields
can be  developed  into $(4 \times
4$)-matrix  $V(x)$ according to $ V(x)= (\; \Psi ^{(1)}, \; \Psi
^{(2)}, \; \Psi ^{(3)},\; \Psi ^{(4)}\; )$; then eqs. $(4.1b)$ read
$$
[\; i \gamma ^{\alpha }(x) \; ( \; \partial _{\alpha} \;  +  \;
 \Gamma _{\alpha }(x)\; ) \; - \; m \; ] \;  V(x) = 0 \; .
\eqno(4.2a)
$$

\noindent Matrix $V(x)$ can be decomposed as
$$
V(x) =  [ \; - i \Phi (x)\; + \; \gamma ^{l} \; \tilde{\Phi
}_{l}(x) \; +  \;
           i \sigma ^{mn} \; \Phi _{mn}(x) \; + \;
   \gamma ^{5} \; \tilde{\Phi }(x) \; + \;  i \gamma ^{l} \gamma ^{5}  \; \tilde{\Phi }_{l}(x)
\;  ]\; E^{-1} \; ; \eqno(4.2b)
$$
\noindent however involved quantities
 $\Phi (x), \; \tilde{\Phi }(x), \; \Phi _{l}(x), \;  \tilde{\Phi }_{l}(x),\;
 \Phi _{mn}(x)$    do not posses   transformation properties of tensor nature with respect to
 the local Lorentz group. In the same time, some quasi-tensor equations can be derived from  $(4.2a)$.
To this end, one should act in the same manner as above. For instance,
turning to  $(2.12c)$
$$
\{ i \gamma ^{c} e^{\beta }_{(c)} \partial_{\beta} \; [\; - i \Psi
+ \gamma ^{l} \Psi _{l} + i \sigma ^{mn} \Psi _{mn} + \gamma ^{5}
\tilde{\Psi} + i \gamma ^{l}\gamma ^{5} \tilde{\Psi }_{l}\; ]\;
E^{-1} \; +
$$
$$
 {i \over2} \gamma _{abc} \gamma^{c} \sigma ^{ab}\; [\; - i \Psi
+ \gamma ^{l} \Psi _{l} + i \sigma ^{mn} \Psi _{mn} + \gamma ^{5}
\tilde{\Psi} + i \gamma ^{l} \gamma ^{5} \tilde{\Psi }_{l}\; ] \;
E^{-1} \; +
$$
$$
 {i \over 2} \gamma _{abc} \gamma ^{c} \; [\; - i \Psi  + \gamma
^{l}\Psi _{l}
 + i \sigma ^{mn} \Psi _{mn} + \gamma ^{5}\tilde{\Psi}
 + i \gamma ^{l}\gamma ^{5} \tilde{\Psi }_{l}\; ] \;
E^{-1} \tilde{\sigma }^{ab}\; -
$$
$$
  m\; [\; - i \Psi  + \gamma ^{l} \Psi _{l} + i \sigma ^{mn} \Psi _{mn}
+ \gamma ^{5} \tilde{\Psi}  + i \gamma ^{l} \gamma ^{5}
\tilde{\Psi }_{l}\; ] \; E^{-1}  \}   = 0  \eqno(4.3)
$$

Multiplying eq.  (4.3)from the left by  $E$ and taking the trace of the result (with the use of
the rule
 $E \; \tilde{\sigma }^{ab} = - \sigma ^{ab} \; E$), which results in
 $$
i \; e^{\beta }_{(c)} \; \mbox{Sp}\;  (\gamma^{c} \gamma^{l})\;
\partial _{\beta} \Psi _{l}
+ {i \over 2}\;  \gamma _{abc} [ \;   \mbox{Sp}\; (\gamma ^{c}
\sigma ^{ab} \gamma ^{l}) \;\Psi _{l} + i \; \mbox{Sp}\; (\gamma
^{c} \sigma ^{ab} \gamma ^{l} \gamma ^{5})\;  \tilde{\Psi }_{l} \;
] +
$$
$$
 {1\over 2} \; \gamma _{abc} [ - \; \mbox{Sp}\; (\gamma^{c} \gamma^{l}
\sigma^{ab}) \; \Psi _{l} - i\; \mbox{Sp}\;  (\gamma^{c}
\gamma^{l} \gamma^{5} \sigma^{ab})\; \tilde{\Psi }_{l}] +  4i \;
\Psi  = 0           \; .
$$

\noindent Further, allowing for the known formulas
$$
{1 \over 2} \; \gamma _{abc} \; \mbox{Sp}\; (\gamma^{c}
\sigma^{ab} \gamma^{l})\; \Psi_{l} = + {1\over 2}\;  \gamma
^{cl}_{\;\;\;\;c} \; ,
$$
$$
-{1 \over 2} \; \gamma_{abc} \; \mbox{Sp}\; (\gamma^{c} \gamma^{l}
\sigma^{ab}) \;
 \Psi _{l} = + {1 \over 2} \; \gamma^{cl}_{\;\;\;\;c} \; ,
$$
$$
 {1 \over 2} \; \gamma _{abc} \; \mbox{Sp}\;  (\gamma^{c} \sigma^{ab} \;
\gamma^{l} \gamma^{5}) \; \tilde{\Psi }_{l} = + {i \over 4}\;
 \gamma _{abc} \epsilon ^{abcl} \;\tilde{\Psi }_{l}  \; ,
$$
$$
-{ 1 \over 2} \; \gamma _{abc} \; \mbox{Sp}\;  (\gamma ^{c} \gamma
^{l} \gamma ^{5} \sigma ^{ab}) \; \tilde{\Psi }_{l} = - {i \over
4}\;
 \gamma _{abc} \; \epsilon ^{abcl}\;  \tilde{\Psi }_{l}     \; .
$$

\noindent for $U$- and $V$-fields respectively we find

\vspace{2mm} for $U$-field
$$
 e^{(l)\alpha} \;
\partial _{\alpha}  \; \Psi _{l} \; + \; \gamma ^{c}_{\;\;lc} \;
\Psi ^{l} \; +  \; m \; \Psi    = 0 \; , \eqno(4.4a)
$$

for $V$-field
$$
 e^{(l)\alpha} \; \partial
_{\alpha}
 \Phi _{l} \; + \; {1 \over 2} \gamma ^{c}_{\;\;lc}(x)\; \Phi ^{l}
-
 {1 \over 4} \gamma_{abc} \;\epsilon
^{abcl} \; \tilde{\Phi}_{l} \;+ \; m\; \Phi    = 0 \; .
\eqno(4.4b)
$$

It should be emphasized that  because  $\Phi$-fields are not of tensor nature under the local Lorentz group
eq.  $(4.4b)$, cannot be presented in pure covariant tensor form  whereas eq.
 $(3.4a)$ does.

In similar manner, acting on eq. (4.3)  by operator  $ \mbox{Sp}\; ( E
\gamma ^{5}\times$ we get

\vspace{2mm} for  $U$-field
$$
 e^{(l)\alpha} \; \partial
_{\alpha } \tilde{\Psi }_{l} \; + \gamma ^{c}_{\;\;lc}\;
\tilde{\Psi }^{l} \; + \; m \; \tilde{\Psi }  = 0 \; ;
\eqno(4.5a)
$$

 for $V$-field
$$
 e^{(l)\alpha } \partial _{\alpha}
 \tilde{\Phi }_{l}  + {1\over 2} \gamma ^{c}_{\;\;lc}
\tilde{\Phi }^{l}  + {1\over 4} \gamma _{abc}\; \epsilon ^{abcl}
\Phi _{l}  + m\; \tilde{\Phi } = 0 \; . \eqno(4.5b)
$$

One more case is when multiplying (4.3)  by  $\;\mbox{Sp}\; (E \gamma ^{k}  $:
$$
i e^{\beta }_{(c)}\; \partial _{\beta}\; [\; -i \;\mbox{Sp}\; (
\gamma ^{k}\gamma ^{c})\; \Psi + i \;\mbox{Sp}\; ( \gamma ^{k}
\gamma ^{c} \sigma ^{mn} ) \Psi _{mn}]\; + \; {i \over 2} \gamma
_{abc} \; [\; -i \;\mbox{Sp}\;( \gamma ^{k}\gamma ^{c}\sigma
^{ab})\; \Psi +
$$

$$
 i \;\mbox{Sp}\; ( \gamma ^{k}\gamma ^{c}\sigma ^{ab}\sigma ^{mn} ) \;  \Psi
_{mn}\; + \;\mbox{Sp}\; ( \gamma ^{k}\gamma ^{c}\sigma ^{ab}\gamma
^{5} )\; \tilde{\Psi } \; ]
 \;+\; {i \over 2} \gamma _{abc}\; [\; +i \;\mbox{Sp}\; (\gamma ^{k} \gamma ^{c} \sigma ^{ab})\;
 \Psi \; -
$$

$$
  i \;\mbox{Sp}\; (\gamma ^{k}\gamma ^{c}\sigma ^{mn}\sigma ^{ab}) \; \Psi
_{mn} \; - \;\mbox{Sp}\; (\gamma ^{k}\gamma ^{c}\gamma ^{5}\sigma
^{ab}) \; \tilde{\Psi } \;] -
 m \;\mbox{Sp}\; (\gamma ^{k} \gamma ^{l}) \; \Psi _{l} = 0 \; ,
$$

\noindent which results in

\vspace{3mm}

for  $U$-field

$$
e^{(k)\alpha} \; \partial _{\alpha} \Psi  \; + \; e^{\alpha
}_{(c)} \;  \partial _{\alpha} \Psi ^{kc}\; + \; \gamma
^{k}_{\;\;mn}  \; \Psi ^{mn} \; + \; \gamma ^{c}_{\;\;lc}
 \; \Psi ^{kl} \; - \;
 m\; \Psi ^{k} = 0 \;  ;
\eqno(4.6a)
$$

\vspace{3mm} for $V$-field

$$
e^{(k)\alpha} \;\partial_{\alpha} \Phi \; + \; e^{\alpha }_{(c)}
\; \partial_{\alpha}\; \Phi ^{kc}\; + \;
 {1\over 2} \gamma ^{k}_{\;\;mn} \Phi ^{mn} \; + \;
{1\over 2} \gamma ^{c}_{\;\;lc} \Phi ^{kl}  \; +
$$

$$
 {1\over 2} \gamma ^{ck}_{\;\;\;\;c}(x) \; \Phi \; +\; {1\over 4}
 \gamma _{abc} \epsilon ^{abck} \; \tilde{\Phi } \; + \;
{1\over 4} \gamma ^{\;\;\;\;k}_{mn} \;\Phi ^{mn} \;-\; m \; \Phi
^{k} = 0     \; . \eqno(4.6b)
$$

\noindent Eqs.  $(4.6a)$ and   $(4.6b)$ substantially differ from each other,
only the first is reduced to  covariant tensor form.
Remaining equations can be treated similarly, the main results are the same:
only for $U$-field there arise covariant tensor equations.

Else one remark about interpretation of the Dirac--K\"{a}hler field  in flat Minkowski space as a set of four Dirac
particles should be given. The matter is that any  particle as a relativistic object
is determined not only by explicitly given wave equation but also determined by
a relevant operation of charge conjugation. The latter, in turn, is fixed by
transformation properties of the wave function  under the Lorentz group.
Evidently, the Dirac--K\"{a}hler object and the system four
Dirac fields assume their own and different charge conjugations.
In particular, having  introduced  a definition for a  particle and antiparticle  in accordance with
four  fermions interpretation, one immediately  see that such a particle-antiparticle separating
turns to be non-invariant with respect to tensor transformation rules of the Dirac-K\"{a}hler constituents.
Thus, even in the flat Minkowski space, the four fermion interpretation  for this field cannot be
evolved with success.

\section{Bosons with different intrinsic parities in curved space-time
}

From the Dirac--K\"{a}hler theory, by imposing special linear restrictions, one can derive more simple
equations  for particle with single  value of spin: ordinary bosons of spin 0 or 1 with different intrinsic parity.

First, let us consider tensor equations in flat Minkowski space with
 four different additional constraints:
$$
\underline{S=0} \qquad \qquad \tilde{\Phi } =0 \; , \qquad
\tilde{\Phi }_{\alpha }=0 \; , \qquad  \Phi _{\alpha \beta } = 0
\; ,
$$
$$
\partial^{l} \Phi_{l} + m \Phi  = 0 \; , \qquad
\partial_{l} \Phi  - m \; \Phi _{l} = 0 \; , \qquad
\partial^{d} \Phi ^{k} -
\partial^{k} \Phi ^{d}  = 0 \;  ;
\eqno(5.1a)
$$

$$
\underline{S = \tilde{0}} \qquad \qquad \Phi  =0 \; , \qquad  \Phi
_{\alpha }=0 \; , \qquad  \Phi _{\alpha \beta } = 0 \; ,
$$
$$
\partial^{l} \tilde{\Phi }_{l} +
 m \; \tilde{\Phi } = 0 \; ,
 \qquad
\partial ^{k } \tilde{\Phi } - m \tilde{\Phi }^{k}  = 0 \; ,
\qquad
\epsilon^{dkcl} \partial_{c} \Phi_{l} = 0 \; ;
\eqno(5.1a)
$$

$$
\underline{S=1} \qquad \qquad \Phi =0 \;  , \qquad \tilde{\Phi}=0
\; , \qquad \tilde{\Phi}_{l}=0 \; ,
$$
$$
\partial^{l} \tilde{\Phi }_{l} +
 m  \tilde{\Phi } = 0 \; ,  \qquad
\partial ^{k } \tilde{\Phi } - m \tilde{\Phi }^{k}  = 0 \; .
\eqno(5.2a)
$$

$$
\underline{S=\tilde{1}:} \qquad \qquad \Phi =0 , \qquad
\tilde{\Phi}=0, \qquad \Phi_{l}=0 \; ,
$$
$$
\partial^{l} \tilde{\Phi}_{l} = 0 \; , \;
\partial^{l} \Phi_{kl} = 0 \; , \;
{1\over 2} \epsilon ^{kcmn}
\partial _{c} \Psi _{mn}  +
m \tilde{\Phi}^{k} = 0 \; , \;
 \epsilon ^{dkcl}
\partial_{c} \tilde{\Phi }_{l} - m \Phi ^{dk} = 0 \; .
\eqno(5.2b)
$$

Let  us describe additional constraints  in spinor form.
For a scalar particle we get
$$
\underline{S = 0 } \;\;   \left | \begin{array}{cc}
\xi  & \Delta  \\
H & \eta  \end{array} \right | = \left | \begin{array}{cc} -
\Phi  \sigma^{2}    & +i  \Phi _{l}  \bar{\sigma}^{l}
\sigma^{2}
\\
-i \Phi _{l}  \sigma^{l}  \sigma^{2}    & + \Phi
\sigma^{2}
\end{array} \right | ,
$$
$$
  \Delta^{tr } = + H\; , \;\; \xi
= - \eta \; , \;\;
      \xi^{tr } = - \xi\; , \;\;  \eta^{tr } = - \eta \; ,
\eqno(5.3)
$$

\noindent the symbol of $tr$ stands for a matrix transposition.
For a pseudoscalar particle we get
$$
\underline{S = \tilde{0}}  \;\;   \left | \begin{array}{cc}
\xi  & \Delta  \\
H & \eta  \end{array} \right | = \left | \begin{array}{cc} +i
\tilde{\Phi}  \sigma^{2}    & -  \tilde{\Phi} _{l}\;
\bar{\sigma}^{l}  \sigma^{2}
\\
- \tilde{\Phi} _{l}  \sigma^{l}  \sigma^{2}    & +i \tilde{\Phi}  \sigma^{2}
\end{array} \right |  ,
$$
$$
  \tilde{\Delta } = - H \; ,\;
\; \xi  = + \eta \; ,  \;\;
 \tilde{\xi } = - \xi  \; , \; \; \tilde{\eta } = - \eta \; .
\eqno(5.4)
$$

\noindent
For a vector particle, we will have
$$
\underline{S = 1}  \qquad   \left | \begin{array}{cc}
\xi  & \Delta  \\
H & \eta  \end{array} \right | = \left | \begin{array}{cc}
 + \; \Sigma^{mn}   \sigma^{2} \Phi_{mn}    &
+i \bar{\sigma}^{l} \sigma^{2}  \Phi _{l}    \\
-i  \sigma^{l} \sigma^{2}    \Phi _{l}     & -
\bar{\Sigma}^{mn}  \sigma^{2}  \Phi_{mn}
\end{array} \right |\; ,
$$
$$
\tilde{\Delta } = + H \; ,
 \;\;  \tilde{\xi } = + \xi \; ,
 \;\;  \tilde{\eta } = + \eta \; .
\eqno(5.5a)
$$

\noindent
Here each of symmetrical spinors $\xi$ and $\eta$  depends on three independent variables:
$$
 \xi + \eta  = -2i \; ( \;  \sigma^{1}
\Phi_{23} + \sigma^{2} \Phi_{31} +
\sigma^{3} \Phi_{12} \; )  \; \sigma^{2} \; ,
$$
$$
\xi - \eta  = 2 \; ( \;  \sigma^{1} \Phi_{01} + \sigma^{2} \Phi_{02} +
\sigma^{3} \Phi_{03} \; )  \; \sigma^{2} \; .
\eqno(5.5b)
$$

\noindent
Finally, a pseudovector case is given by
$$
\underline{S = \tilde{1}} \qquad  \qquad  \left | \begin{array}{cc}
\xi  & \Delta  \\
H & \eta  \end{array} \right | = \left | \begin{array}{cc}
 + \Sigma^{mn} \sigma^{2} \Phi_{mn}      &
- \bar{\sigma}^{l} \sigma^{2}   \tilde{\Phi}_{l}  \\
- \sigma^{l}\sigma^{2}    \tilde{\Phi}_{l}   &
 - \bar{\Sigma}^{mn} \sigma^{2} \Phi_{mn}
\end{array} \right | \; ,
$$
$$
 \tilde{\Delta } = - H \;
, \; \; \tilde{\xi } = + \xi\; ,   \; \; \tilde{\eta } = + \eta \; .
\eqno(5.6)
$$

We are to extend this approach to general covariant case
$$
\nabla ^{\alpha } \Psi _{\alpha }    + m \Psi  = 0 \; , \qquad
\tilde{\nabla }^{\alpha } \Psi _{l} + m \tilde{\Psi}   = 0
\;  ,
$$
$$
\nabla _{\alpha }\Psi  + \nabla ^{\beta } \Psi _{\alpha \beta
}(x) -
 m \Psi _{\alpha }= 0 \; ,
$$
$$
 \tilde{\nabla }_{\alpha }\Psi   -
{1\over 2} \epsilon ^{\;\;\beta \rho \sigma }_{\alpha }(x)
 \nabla _{\beta } \Psi _{\rho \sigma } -
m \tilde{\Psi }_{\alpha }  = 0 \; ,
$$
$$
 \nabla _{\alpha } \Psi _{\beta } -
\nabla _{\beta } \Psi _{\alpha } + \epsilon ^{\;\;\;\;\rho
\sigma }_{\alpha \beta }(x) \nabla _{\rho }
 \tilde{\Psi }_{\sigma }  - m \Psi _{\alpha \beta } = 0 \; .
\eqno(5.7)
$$

First let it be
$$
\underline{S=0,} \qquad
\nabla ^{\alpha } \Psi_{\alpha } + m \Psi  = 0 \; ,
$$
$$
\nabla _{\alpha } \Psi  - m \Psi _{\alpha } = 0 \; ,
\;
\nabla_{\alpha} \Psi_{\beta} - \nabla_{\beta} \Psi_{\alpha}
= 0 \;  ,
\eqno(5.8)
$$

\noindent
two first  are  the Proca  equations for scalar particle, the last  equation holds identically
$$
\partial _{\alpha } \; \partial _{\beta }\; \Psi  \; - \;
\Gamma ^{\mu }_{\alpha \beta }\; \partial _{\mu } \; \Psi
\; -  \;  \partial _{\beta} \; \partial _{\alpha } \; \Psi \;
+ \; \Gamma ^{\mu }_{\beta \alpha }  \; \partial _{\mu }\;
\Psi   = 0 \; .
$$

For a pseudoscalar field we have
$$
\underline{S= \tilde{0} \; ,}\;\;
\nabla ^{\alpha } \tilde{\Psi }_{\alpha } +
 m \tilde{\Psi }  = 0 \; ,
 $$
 $$
\nabla _{\alpha } \tilde{\Psi } - m \tilde{\Psi }_{\alpha }
= 0 \; ,
\;
  \epsilon_{\alpha \beta}^{\;\;\;\;\rho \sigma} (x) \nabla_{\rho} \Psi_{\sigma} = 0 \; ;
\eqno(5.9)
$$

\noindent here the last equation holds identically.
Now, let  only $\Psi _{\alpha } \neq  0 , \; \Psi
_{\alpha \beta }(x) \neq  0$, then
$$
\underline{S=1\; ,}  \qquad\qquad \nabla^{\alpha} \Psi_{\alpha} = 0 \; , \; \nabla ^{\beta }
\Psi _{\alpha \beta } - m \Psi _{\alpha }  =  0 \;  ,
$$
$$
-{1\over 2} \epsilon_{\alpha}^{\;\;\beta \rho \sigma}(x)
\nabla_{\beta} \Psi_{\rho \sigma} = 0 \; ,
\;
\nabla_{\alpha} \Psi_{\beta} - \nabla_{\beta}\Psi_{\alpha }
= m \Psi _{\alpha \beta }  \; .
\eqno(5.10)
$$

\noindent Here the first and third equation hold identically:
$$
\nabla ^{\alpha } \Psi _{\alpha } = {1\over m} \; \nabla
^{\alpha } \nabla ^{\beta }  \; \Psi _{\alpha \beta } = {1\over
2m}\; [ \; \Psi _{\alpha \nu } \; R^{\nu \;\; \beta \alpha
}_{\;\;\beta }
$$
$$
- \Psi _{\beta \nu } R^{\nu \;\;  \beta \alpha }_{\;\;\alpha
} \; ] = {1\over 2m} \; [\; - \Psi _{\alpha \nu } \; R^{\nu
\alpha } - \Psi _{\beta \nu } \; R^{\nu \beta }\;  ] = 0 \; ,
$$

$$
-{ 1\over 2m} \; \epsilon ^{\;\;\beta \rho \sigma }_{\alpha }(x)
\;
 \nabla _{\beta } \; [ \; \nabla _{\rho } \Psi _{\sigma } -
                    \nabla _{\sigma } \Psi _{\rho } \; ]
$$
$$
= -{1\over 4m} \; \epsilon ^{\;\;\rho \beta \sigma }_{\alpha }(x) \;
[ \; ( \nabla _{\beta } \nabla _{\rho } -
    \nabla _{\rho } \nabla _{\beta } ) \; \Psi _{\sigma } -
    ( \nabla _{\beta } \nabla _{\sigma } -
    \nabla _{\sigma } \nabla _{\beta } )\;  \Psi _{\rho } \; ]
$$
$$
= -{1\over 4m} \;  \epsilon ^{\;\;\beta \rho \sigma }_{\alpha
}(x)\;
 ( \Psi ^{\nu } \; R_{\nu \sigma \rho \beta } -
   \Psi ^{\nu } \; R_{\nu \rho \sigma \beta } ) \; = 0 \; .
$$

Now, let  $\Psi (x) =
 \tilde{\Psi } = \Psi _{\alpha } = 0$, then
$$
\underline{S= \tilde{1}}\;, \qquad \qquad \qquad
\nabla^{\alpha} \tilde{\Psi}_{\alpha} = 0 \; , \;
\nabla^{\beta} \tilde{\Psi}_{\alpha \beta} = 0 \; ,
$$
$$
{1\over 2}\; \epsilon ^{\;\;\beta \rho \sigma }_{\alpha }(x)\;
\nabla _{\beta } \Psi _{\rho \sigma } + m \;\tilde{\Psi } =
0 \; ,
$$
$$
 \epsilon ^{\;\;\;\;\rho \sigma }_{\alpha \beta }(x) \;
\nabla _{\rho } \tilde{\Psi }_{\sigma } - m \;\Psi _{\alpha
\beta } = 0 \; .
\eqno(5.11)
$$

\noindent The first and the second equations  hold identically:
$$
\nabla ^{\alpha } \tilde{\Psi }_{\alpha } = -  {1\over 2m}\;
\nabla ^{\alpha } \epsilon ^{\;\;\beta \rho \sigma }_{\alpha }(x)
\; \nabla _{\beta } \; \Psi _{\rho \sigma }  =
 -{1\over 2m}\; \epsilon ^{\;\;\beta \rho \sigma }_{\alpha }(x)
\; \nabla ^{\alpha } \nabla _{\beta } \; \Psi _{\rho \sigma }
$$
$$
= -{1\over 4m} \epsilon ^{\;\;\beta \rho \sigma }_{\alpha }(x) [\;
\Psi _{\nu \sigma }\;  R^{\nu \;\;\;\; \alpha }_{\;\;\beta \rho
}(x) \; + \; \Psi _{\rho \nu } \; R^{\nu \;\;\;\; \alpha
}_{\;\;\sigma \beta } \; ] \; ,
$$

$$
\nabla ^{\beta } \; \Psi _{\alpha \beta }(x) = {1\over m } \;
\nabla ^{\beta } \; \epsilon ^{\;\;\;\;\rho \sigma }_{\alpha
\beta }(x)\; \nabla _{\rho } \Psi _{\sigma }  =
 {1\over 2m}  \; \epsilon ^{\;\;\rho \sigma }_{\alpha \beta }(x) \;
 \Psi ^{\nu } \; R_{\nu \sigma \rho \beta } \; .
$$

Constraints separating four boson fields are the same as in the case of Minkowski space:

$$
S = 0 \; ,\qquad  \qquad  \tilde{\Delta } = + H\; , \;
\xi  = - \eta \; , \;
      \tilde{\xi } = - \xi\; , \;  \tilde{\eta } = - \eta \; .
$$
$$
S = \tilde{0} \; ,\qquad  \qquad  \tilde{\Delta } = - H
\; , \; \xi  = + \eta \; ,  \;
 \tilde{\xi } = - \xi  \; , \; \tilde{\eta } = - \eta \; .
$$
$$
S = 1 \; , \qquad  \qquad \tilde{\Delta } = + H \; ,
 \; \tilde{\xi } = + \xi \; ,
 \; \tilde{\eta } = + \eta \; .
$$
$$
S = \tilde{1} \; , \qquad \qquad \tilde{\Delta } = - H
\; , \; \tilde{\xi } = + \xi\; ,   \; \tilde{\eta } = + \eta \; .
\eqno(5.12)
$$

\section{Discussion
}

We may conclude that the use of tetrad formalism permit us to
apply results on classification of the particles  with respect to discrete Lorentzian transformations
(including intrinsic parity of bosons)
when treating relevant particle fields on the background of arbitrary
 curved space-time model.


\begin{thebibliography}{9999}


\bibitem{1939-Wigner}
  Wigner  E.P.
On unitary representations of the  inhomogeneous Lorentz group.
 Ann. of Math.  1939. Vol. 40. P. 149 -- 204.


\bibitem{1941-Pauli}
  Pauli W.
 Relativistic field theories of elementary particles.  Rev. Mod. Phys. 1941. Vol. 13. P. 203 -- 232.


\bibitem{1949-Bhabha}
  Bhabha H.J.  On the postulational basis of the theory of
elementary particles. Rev. Mod. Phys. 1949. Vol. 21,  no 3. P. 451 -- 462.


\bibitem{1947-Yarish-Chandra}
 Harish-Chandra. On relativistic wave equation. Phys. Rev. 1947. Vol.  71,  no 11. P.  793 -- 805.


\bibitem{1948-Gel'fand}
Gel'fand I.M., Yaglom A.M.
 General relativistically invariant equation and infinite
 infinite-dimensional, representations of the Lorent group
 JETP. 1948. Vol.  18, no 8.  P. 703 -- 733.


\bibitem{1953-Corson}
Corson E.M.
 Introduction to tensors, spinors and relativistic
wave equations. London: Blackie and Son,  1953.


\bibitem{1956-Umezawa}
 Umezawa H.
 Quantum Field Theory. Amsterdam (North-Holland),    1956.


\bibitem{1957-Shirokov}
  Yu.M. Shirokov.
  Group theory consideration of the bases
of relativistic quantum mechanics I.
 JETP. 1957. Vol. 33. No 4.  P. 861-872;
II.   JETP. 1957. Vol.  33. No 5.  P. 1196-1207; III.   JETF. 1957.
Vol. 33.  No 5.  P. 1208-1214; IV.   JETP. 1958. Vol. 34. No 3.  P.


\bibitem{1968-Bogush-Moroz}
Bogush A.A., Moroz L.G. Introduction to klassical field theory.
Minsk, Nauka i tekhnika,  1968.

\bibitem{1979-Fedorov}
Fedorov F,I, The Lorentz group. Moskow, 1979.


\bibitem{1928-Ivanenko}
Ivanenko D.,    Landau L.
Zur theorie des magnetischen  electrons. Zeit. Phys.  1928.  Bd. 48,  N  8. S. 340 -- 348.


\bibitem{1929-Lanczos(1)}
  Lanczos C.
The tensor analytical relationships of Dirac's
equation. Zeit.  Phys.  1929. Bd. 57. S. 447 -- 473.


\bibitem{1929-Lanczos(2)}
  Lanczos C.
The covariant formulation of Dirac's equation.
 Zeit. Phys.  1929. Bd. 57. S.  474 -- 483;
The conservation laws in the field theoretical representation of
Dirac's theory.
   Zeit.  Phys.  1929. Bd. 57. S. 484 -- 493.


\bibitem{1930-Juvet}
  Juvet G.
 Op\'{e}rateurs de Dirac et \'{e}quations de Maxwell. Comm. Math. Helv. 1930. Vol. 2. P.  225 -- 235.


\bibitem{1932-Juvet}
 Juvet G.,  Schidlof A.
 Sur les nombres hypercomplexes de Clifford
et leurs applications  $\grave{a}$  l'analyse vectorielle
ordinaire, $\grave{a}$  l'\'{e}lectromagn\'{e}tisme de Minkowski
et $\grave{a}$ la th\'{e}orie de Dirac.
 Bull. Soc. Sci. Nat. Neuch$\hat{a}$tel.  1932. Vol.  57. P.  127 -- 141.


\bibitem{1932-Einstein(1)}
Einstein A.,   Mayer V. Semivektoren   und   Spinoren.
Sitz. Ber. Preuss. Akad. Wiss. Berlin. Phys.-Math. Kl.
1932. S. 522 -- 550.


\bibitem{1933-Einstein(1)}
Einstein A.,    Mayer W.
Die Diracgleichungen f\"{u}r
Semivektoren.
Proc. Akad.  Wet. Amsterdam. 1933. Bd. 36. S. 497 -- 516.




\bibitem{1933-Einstein(2)}
  Einstein A.,   Mayer W. Spaltung der Nat\"urlichsten
Feldgleichungen  f\"ur Semi-Vektoren in  Spinor-Gleichungen  von
Diracschen Tipus. Proc. Akad. Wet. Amsterdam. 1933. Bd. 36. S. 615 -- 619.

\bibitem{1934-Einstein}
 Einstein A.,   Mayer W.
Darstellung  der  Semi-Vektoren  als
gew\"{o}hnliche Vektoren von Besonderem  Differentiations
Charakter.  Ann. of  Math. 1934. Vol. 35,  N  1.  P. 104 -- 110.



\bibitem{1935-Frenkel}
Frenkel Ya.I. Electrodynamics.  Vol. I,  1934;  Vol. II,  1935.


\bibitem{1937-Whittaker}
  Whittaker E.T.
 On the relations of the tensor-calculus to the spinor-calculus.
  Proc. Roy. Soc. London. A. 1937.  Vol. 158. P. 38 -- 46.



%133
\bibitem{1937-Proca}
  Proca A.
Sur  un  article  de M.E. Whittaker,  intitul\'e  "Les
relations  entre le  calcul tensoriel  et le  calcul   des
spineurs".
 J. Phys. et Radium. 1937. Vol.   8. P. 363 -- 365.


\bibitem{1936-Ruse}
   Ruse  H.S.
 On the geometry of Dirac's equations and their
expression in tensor form.
 Proc. Roy. Soc. Edin.   1936. Vol. 57. P. 97 -- 127.

\bibitem{1939-Taub(1)}
  Taub A.H.
 Tensors equations equivalent to the Dirac equations.
 Ann. Math. 1939. Vol. 40. P. 937.



\bibitem{1939-Taub(2)}
   Taub A.H.
Spinor equations for the meson and their solution when no field is
present. Phys. Rev. 1939. Vol. 56,  N 8. P. 799 -- 810.

\bibitem{1939-Belinfante(1)}
  Belinfante F.J.
The undor equation of the meson field. Physica. 1939. Vol. 6. P. 870.






\bibitem{1939-Belinfante(2)}
  Belinfante  F.J.
Spin of Mesons. Physica. 1939. Vol. 6. P. 887 -- 898.


\bibitem{1951-Ivanenko}
Ivanenko D.,  Sokolov A. Quantim field theory.
Moscov,  1951.


\bibitem{1958-Feshbach(2)}
  Feshbach H.,  Nickols W.
 A wave equation for a particle of maximum spin one.
Ann. Phys. N.Y. 1958. Vol. 4,  N  4. P. 448 -- 458.


\bibitem{1960-Kahler}
 K\"{a}hler E.
Innerer and \"{a}usserer Differentialkalk\"{u}l.
 Abh. Dt. Akad. Wiss. Berlin. Kl. Math.-Phys. u. Techn. 1960.  N 4.


\bibitem{1961-Kahler}
 K\"ahler E.
 Die Dirac-Gleichunung.
Abh. Dt. Akad. Wiss. Berlin, Kl. Math.-Phys. u. Techn.
1961,  N 1.


\bibitem{1962-Leutwyler}
 Leutwyler H.
Generally covariant Dirac equation and associated boson fields.
// Nuovo Cimento. 1962. Vol. 26,  N 5. P. 1066.


\bibitem{1964-Klauder}
 Klauder J.R.
 Linear representation of spinor fields by antysymmetric tensors.
 J. Math. Phys. 1964. Vol. 5,  N  9. P. 1204 -- 1214.

\bibitem{1965-Penney}
 Penney R. Tensorial description of neutrinos.
 J. Math. Phys. 1965. Vol. 6,  N  7. P. 1026 -- 1028.


\bibitem{1967-Cereignani}
 Cereignani C.
 Linear representations of spinors by tensors.
J. Math. Phys. 1967. Vol. 8,  N 3. P. 417 -- 422.


\bibitem{1970-Streater}
 Streater  R.F.,   Wilde I.F.
 Fermion states of a boson field.
Nucl. Phys. B. 1970. Vol. 24.  P. 561.


\bibitem{1971-Pestov}
Pestov A.B. Connection betwee Dirac and Maxwell equations.
Dubna, 1971. 18 pages (Prepeint  P2-5798).

\bibitem{1978-Pestov}
Pestov A.B.
Reltivistic equations defined by exterior derivative operators and extended divergence.
    TMP. 1978. Vol. 34, no  1. P. 48 -- 57.



\bibitem{1983-Pestov}
Pestov A.B.
On the group of internal  symmetry  of  the wave equation
defined by exterior derivative operators. Dubna, 1983.  (Preprin P2-83-506).


\bibitem{1973-Osterwalder}
  Osterwalder K.
 Duality for free Bose fields.
 Commun. Math. Phys. 1973. Vol. 29,  N 1. P. 1 -- 14.


\bibitem{1975-Crumeyrolle}
 Crumeyrolle A.
 Une th\'{e}orie de Einstein -- Dirac en spin maximum 1.  Ann. Inst. H. Poincar\'{e}.   A. 1975.   Vol.   22. P.  43.






\bibitem{1975-Durand}
  Durand E.
 16-component theory  of the spin-1 particle  and its
generalization to arbitrary spin.  Phys. Rev. D. 1975.  Vol. 11,  N  12. P. 3405 -- 3416.

\bibitem{1977-Strazhev}
Strazhev V.I. On the symmetry group of extended equations for a vector
field. Izvestiya Vuzov. Fizika. 1977, no  8. P. 45 -- 48.


\bibitem{1978-Kruglov}
 Kruglov S.I., Strazhev V.I.
 Internal symmetries and conservation lows in ckassical theory of a vector field of general type.
 Izvestiya Vuzov. Fizika.
 1978, no 4. P. 77 -- 81.


\bibitem{1978-Strazhev(1)}
Strazhev V.I.
On dyad symmetry of a vector filed of general type.
 Acta Phys. Pol. B. 1978. Vol. 9  P. 449 -- 458.

\bibitem{1978-Bogush(1)}
Bogush A.A., Kruglov S.I., Strazhev V.I.
 On the group of internal symmetry of 16-component
 neory of a cector particles.
 Doklady AN BSSR. 1978. Vol.  22,  no 10.    P. 893 -- 895.




\bibitem{1978-Bogush(2)}
 Bogush A.A., Kruglov S.I.
 On equation of vector field of general type.
 Procceding of Academy of Sciences of BSSR. ser. phys.-mat. 1978, no  4.   P. 58 -- 65.


\bibitem{1987-Satikov}
Satikov I.A., Strazhev V.I.
On quantum description of the Dirac--K\"{a}hler field.
TMP. 1987.  Vol.  73,  no 1. P. 16 -- 25.


\bibitem{1988-Strazhev(1)}
Strazhev V.I., Pletjuxov, Fedorov F.I.
On connection of spin and statistics in the theory
of relativistic wave equations  with intrinsic degrees of freedom.
Minsk,    1988. 36 pages. (Preprint no 517 /  IP AN BSSR).






\bibitem{1988-Strazhev(2)}
 Strazhev V.I.,   Berezin A.V., Satikov I.A.
 Dirac--K\"{a}hler equations  and quantum theory of the Dirac field
 witn internal symmetry group
$SU(2,2)$. Minsk,   1988. 20 pages . (Preprint no   522 / IP AN BSSR).


\bibitem{2002-Strazhev}
 Strazhev V.I., Tsionenko D.A.
  On Dirac--K\"{a}hler gauge field teory in a curved space-time.
  Vestnik BGU. ser. I, fiz.-mat.-inform. с.  2002, no  2.  P. 15 -- 21.


\bibitem{2002-Tzionenko}
Tsionenko D.A.
Уравнение Дирака-Кэлера как координатное представление
квантовомеханического уравнения движения
// Весцi НАН Беларусi. Сер. фiз.-мат. навук. 2002, no 4. С. 75 -- 83.


\bibitem{2003-Tzionenko}
Tsionenko D.A.
Dirac--K\"{a}hler equation in non-Euclidean space-time.
Procceding of National Academy of Sciences of Belarus. phys.-mat. 2003, no 1. P. 81 -- 85.


\bibitem{2007-Strazhev}
Strazhev V.I.,    Satikov I.A.,  Tsionenko D.A.
Dirac--K\"{a}hler equation, classical theory. Minsk, BGU,  2007.


\bibitem{1978-Graf}
  Graf W.
 Differential forms as spinors.  Ann. Inst.  H. Poincar\'e. A. 1978. Vol.  29, N  1. P. 85 -- 109.


\bibitem{1982-Benn}
 Benn  I.M.,   Tucker R.W.
 A generation  model based on K\"{a}hler fermions.
Phys. Lett. B. 1982. Vol. 119,  N 4-6.  P. 348 -- 350.


\bibitem{1983-Benn(1)}
  Benn  I.M.,  Tucker R.W.
Fermions without spinors.
 Commun.   Math. Phys. 1983. Vol. 89,  N  3. P. 341 -- 362.







\bibitem{1983-Benn(2)}
   Benn I.M.,   Tucker R.W.
 K\"{a}hler  fields and five-dimensional Kaluza -- Klein theory.
 J. Phys. A. 1983. Vol. 16,  N 4.  P. 123 -- 125.





\bibitem{1983-Benn(3)}
 Benn  I.M.,    Tucker R.W.
 Clifford analysis of exterior forms and
Fermi-Bose symmetry.
J. Phys. A. 1983.  Vol. 16,  N 17. P. 4147 -- 4153.


\bibitem{1983-Benn}
  Benn  I.M.,  Tucker R.W.
 A local right-spin covariant K\"{a}hler  equation.
Phys. Lett. B. 1983. Vol. 130,  N 3-4. P. 177 -- 178.



\bibitem{1985-Benn}
Tucker  R.W.,   Benn  I.M.
 The differential   approach to spinors and their symmetries.
 Nuovo Cim. A. 1985. Vol. 88. Ser. 2,  N 3. P. 273 -- 285.

\bibitem{1982-Banks}
 Banks T.,  Dothan Y.,  Horn  D.
Geometric fermions. Phys. Lett. B. 1982. Vol. 117,  N  6. P. 413 -- 417.




\bibitem{1982-Garbaczewski}
 Garbaczewski P.
Quantization  of spinor fields.  Meaning of
"bosonization" \hspace{1mm}  in (1+1) and (1+3) dimensions.
J. Math. Phys.  1982. Vol. 23, N 3. P. 442 -- 450.





\bibitem{1982-Pletjuxov}
Pletjuxov V.A., Strazhev V.I.
On Dirac like wave equation for particles with maxmal spin 1
Doklady AN BSSR. 1982. Vol.  26, no  8. P.  691 -- 693.



\bibitem{1986-Pletjuxov}
Pletjuxov V.A., Satikov I.A., Strazhev V.I.
Relativistic wave  equations and massless Dirac--K\"{a}hler  field.
Covariant methods in theoretical physics. Elementary particle physica and relativity theoory.
Minsk, Institute of Physics, 1986. P. 31 -- 35.





\bibitem{1987-Pletjuxov}
Pletjuxov V.A.,  Strazhev V.I.
 On possible extensions of the Dirac--K\"{a}hler  field
Vesti AN BSSR. ser. fiz.-mat.
 1987,  no 5.  P. 87 -- 92.


\bibitem{1989-Pletjuxov}
Pletjuxov V.A.,  Strazhev V.I.
Tensorial equations and Dirac particles  with internal
 degrees of freedom.
 Yadernaya Fizika.  1989. Vol. 49. P. 1505 -- 1514.



\bibitem{1983-Holland}
 Holland   P.R.
  Tensor conditions for algebraic  spinors.
J. Phys. A.  1983. Vol. 16,  N  11. P. 2363 -- 2374.

\bibitem{1985-Ivanenko}
 Ivanenko D.D.,  Obukhov Yu.N.,  Solodukhin  S.N.
On antisymmetric tensor representation.
of the Dirac equation.  Trieste, 1985 (Preprint IC/85/2. ICTP).


\bibitem{1993-Obukhov}
 Obukhov Yu.N.,  Solodukhin  S.N.
 Reduction of the Dirac equation and its coonection to Ivanenko--Landau-- K\"{a}hler equation.
TMP.  1993. Vol.  Vol. 94. P. 276 -- 295.


\bibitem{1986-Bullinaria}
  Bullinaria J.A.
  K\"{a}hler fermions in arbitrary space-times,
their dimensional reduction and relation to spinorial fermions.
Ann. Phys. (N.Y.). 1986.  Vol. 168,  N 2. P. 301 -- 343.


\bibitem{1986-Blau}
 Blau M.
 Clifford algebras and K\"{a}hler -- Dirac spinors.
 Ph.D. dissertation, Report UWTH Ph 198616. Universitat Wien,  1986. 200 p.


\bibitem{1987-Jourjine}
 Jourjine   A.N.
  Space-time Dirac -- K\"{a}hler  spinors.  Phys. Rev. D.   1987. Vol. 35,  N 2. P.  757 -- 758.


\bibitem{1989-Krolikowski(1)}
  Krolikowski W.
Dirac equation with hidden extra spin: a
generalization of k\"{a}hler equation. I.
Acta Phys. Polon. B. 1989.  Vol. 20,  no 10. P. 849 -- 858.



\bibitem{1989-Krolikowski(2)}
Krolikowski W.
 Dirac equation with hidden extra spin: a
generalization of K\"{a}hler equation. II.
Acta Phys. Polon. B. 1990.  Vol. 21,  no  3. P. 201 -- 207.


\bibitem{1989-Howe}
 Howe P.
A particle mechanics description of antisymmetric tensor
fields.  Class. Quant. Grav. 1989. Vol. 6.  P. 1125.


\bibitem{1995-Nikitin(1)}
  Beckers J.,   Debergh N.,   Nikitin A.G.
On parasupersymmetries and
relativistic descriptions for spin one particles. I. The free
context.
Fortschr.  Phys. 1995. Vol. 43,  N 1. P. 67 -- 80;
 II. The interacting context (with electromagnetic fields)
 Fortschr.  Phys. 1995. Vol. 43,  N 1. P. 81 -- 96.


\bibitem{1978-Bogush(2)}
 Bogush A.A., Kruglov S.I.
On equations for a vectot fiel of general type.
Vesti AN BSSR. ser. fiz.-mat.  1978, no   4.   P. 58 -- 65.


\bibitem{2000-Kruglov}
  Kruglov S.I.
Symmetry and electromagnetic interactions of Fields with multispin.  N.Y.:
Nova Science Pub. Inc., Hauppauge, 2000.

\bibitem{2002-Kruglov}
  Kruglov S.I. Dirac -- K\"ahler equations.  Intern. J. Theor. Phys. 2002. Vol. 41. P. 653 -- 687.

\bibitem{2004-Kruglov(1)}
  Kruglov S.I.
 On the generalized Dirac equation for fermions with two mass states.
Ann.  Fond. L. de Broglie. 2004. Vol.  29,
Hors s\'erie 2. P.  1005 -- 1016.





\bibitem{1998-Marchuk}
  Marchuk N.G.
Dirac gamma-equation, classical gauge fields and
Clifford algebra.
 Adv. Appl.Clifford Alg. 1998. Vol.  8. P.  181 -- 2242.


\bibitem{1999-Marchuk(1)}
 Marchuk N.G.
  Gauge fields of the matrix Dirac equation.
Nuovo Cim. B.  1998. Vol. 113. P. 1287 -- 1295.





\bibitem{1999-Marchuk(2)}
 Marchuk N.G.
 A gauge model with spinor group for a description of local interaction of a fermion
 with electromagnetic and gravitational fields.  Nuovo Cim. B. 2000.  Vol. 115. P.  11 -- 25.


\bibitem{2001-Marchuk}
 Marchuk N.G.
 A tensor form of the Dirac equation.
 Nuovo Cim. B. 2001. Vol. 116, N  10.  P. 1225 -- 1248.


\bibitem{2002-Marchuk(1)}
 Marchuk N.G.
 Dirac-type tensor equations with non-Abelian gauge symmetries on pseudo-Riemannian space.
Nuovo Cim. B. 2002. Vol.117. P. 95 -- 120.




\bibitem{2002-Marchuk(2)}
Marchuk N.G.
 The Dirac equation vs. the Dirac type tensor equation.
Nuovo Cim. B. 2002. Vol. 117. P. 511 -- 520.




\bibitem{2002-Marchuk(3)}
Marchuk N.
 A concept of Dirac-type tensor equations. arXiv:math-ph/0212006.


\bibitem{2005-Krivskij}
Krivskij I.Yu., Lompej R.R. Simulik B.M.
On symmetries of complex  Dirac -- K\"ahler equation.
TMP. 2005. Vol.  143. P. 64 -- 82.


\bibitem{1973-Landau}
Landau L.D., Lifshitz E.M.
Theoretical physics, II. Field theory.
Moscow, 1973.



\end{thebibliography}
\end{document}